\documentclass[apj]{emulateapj}

\newcommand{\um}{\,\micron}

\newcommand{\arIII}{[Ar\,\textsc{iii}]}
\newcommand{\neII}{[Ne\,\textsc{ii}]}

\newcommand{\sIV}{[S\,\textsc{iv}]}

\newcommand{\cbsm}{\textsc{Cubism}}

\shortauthors{Smith \textit{et al.}}  
\shorttitle{Spectral Mapping Reconstruction}

\begin{document}

\title{Spectral Mapping Reconstruction of Extended Sources}

\author{%
J.D.T. Smith\altaffilmark{1}, 
L. Armus\altaffilmark{2}, 
D.A. Dale\altaffilmark{3}, 
H. Roussel\altaffilmark{4}, 
K. Sheth\altaffilmark{2}, 
B.A. Buckalew\altaffilmark{5},
T.H. Jarrett\altaffilmark{5},
G. Helou\altaffilmark{5}, and  
R.C. Kennicutt, Jr.\altaffilmark{6,1}
}

\altaffiltext{1}{Steward Observatory, University of Arizona, Tucson, AZ
  85721} 
\altaffiltext{2}{Spitzer Science Center, Caltech, Pasadena, CA}
\altaffiltext{3}{Dept. of Physics \&
  Astron., Univ. of Wyoming, Laramie, WY}
\altaffiltext{4}{Max-Planck-Institut f\"{u}r Astronomie, Heidelberg, Germany}
\altaffiltext{5}{Caltech, Pasadena, CA} 
\altaffiltext{6}{Institute of Astronomy, Cambridge University, Cambridge, UK}

\email{jdsmith@as.arizona.edu}

\begin{abstract}
  Three dimensional spectroscopy of extended sources is typically
  performed with dedicated integral field spectrographs.  We describe a
  method of reconstructing full spectral cubes, with two spatial and one
  spectral dimension, from rastered spectral mapping observations
  employing a single slit in a traditional slit spectrograph.  When the
  background and image characteristics are stable, as is often achieved
  in space, the use of traditional long slits for integral field
  spectroscopy can substantially reduce instrument complexity over
  dedicated integral field designs, without loss of mapping efficiency
  --- particularly compelling when a long slit mode for single
  unresolved source followup is separately required.  We detail a custom
  flux-conserving cube reconstruction algorithm, discuss issues of
  extended source flux calibration, and describe \cbsm, a tool which
  implements these methods for spectral maps obtained with ther Spitzer
  Space Telescope's Infrared Spectrograph.

\end{abstract}


\section{Introduction}

Spectroscopy has long been at the forefront of astronomical research.
From uncovering the elemental abundances underlying the solar spectrum,
to modern massively parallel surveys probing the scale and structure of
the universe, the dispersion of light affords an almost preternatural
power of discovery at a distance.  The recent advent of large format
optical and infrared detector arrays has revolutionized spectroscopy,
enabling many novel methods which capitalize on the growing pixel count.
Cross-dispersion of the dispersed beam enables higher spectral
resolution and longer slits in compact instruments
\citep[e.g.][]{Allington-Smith1989,Tull1995,1998SPIE.3354..798S,
  McLean1998,2001PASP..113..227W}.  The use of multiple slits or fibers
makes it possible to obtain spectra for hundreds of objects
simultaneously over wide fields
\citep[e.g.][]{Heacox1992,York2000,Colless2001,LeFevre2001,Hook2003}.
Integral field spectroscopy of moderately wide areas using image
slicing, densely packed fiber bundles, or micro-lens arrays efficiently
map extended objects \citep[e.g.][]{Tecza1998,Content1998,Mandel2000,
  Dubbeldam2000,Bacon2001,Poglitsch2006}.  These methods, along with
Fabry-Perot scanning and other imaging interferometers, fall under the
broad category of \textit{three dimensional} spectroscopy, which
encompasses a set of techniques and instrumental designs which enable
the observer to build up spectral cubes with two spatial and one
spectral dimension.

Given finite detector area, a tradeoff among field of view, spectral
resolution, and instantaneous wavelength coverage must be made for a
given three dimensional spectrograph design.  In the limit of zero
background, identical throughput and detector performance, and an
unchanging point spread function (PSF), any combination of these three
which utilizes the same fractional detector area will offer identical
spectral mapping efficiency.  Modern integral field spectrographs
typically favor larger fields at the cost of resolution or wavelength
coverage, but have the notable advantage of capturing the entire field
at once, which mitigates systematics due to variations in sky background
and transmission, and time variations in the PSF \citep[e.g. in adaptive
optics applications,][]{Thatte2000}.  For observing single unresolved
sources, however, integral field spectrographs are relatively
inefficient, since they dedicate a large fraction of their detector
budget to unused background area.  For isolated point sources, single
slit spectroscopy still maintains an advantage.  This advantage is even
more acute in the ground-based infrared, where, due to the large and
variable sky brightness, source and sky must be switched rapidly
throughout the observation --- on timescales of minutes at near infrared
(NIR) wavelengths (1--2.5\um) and up to 10Hz at thermal mid-infrared
(MIR) wavelengths (4--20\um).  In this regime, single long slits offer
another compelling advantage: the ability to chop the source between two
positions along the slit, performing \emph{in situ} sky subtraction with
well-characterized slit throughput, and minimal loss of observing
efficiency.

Unfortunately, single long slits are of more limited use for three
dimensional spectroscopy of extended sources from the ground.  In
particular, variations in the sky brightness and transmission, as well
as the PSF during the (potentially long) series of integrations required
to map out large extended regions pose significant challenges,
especially for sources with surface brightness comparable to or below
that of the sky.  Despite these drawbacks, the scanning method has been
used at optical and near-infrared wavelengths to construct
moderate-sized maps of bright sources
\citep[e.g.][]{Burton1992,Monteiro2005}, and in solar astronomy
\citep[e.g.][]{Johanneson1992}, but has been eclipsed in recent years by
dedicated integral field techniques.  Though they offer high mapping
efficiency for extended sources, the reduced efficiency of integral
field spectrographs for unresolved sources often dictates that
additional long-slit modes are included in instrument which offer these
capabilities, increasing complexity.

In space, the various limitations of long slits for conducting three
dimensional spectroscopy over wide fields are minimized, since the
astrophysical background is typically smooth and unvarying, and the
image characteristics fixed.  The Infrared Spectrograph
\citep[IRS,][]{Houck2004a} aboard the Spitzer Space Telescope
\citep{Werner2004} includes four sensitive, background-limited
spectrograph modules (covering 5--38\um) based on fixed single slits,
with no moving parts.  The Spitzer telescope offers a dedicated spectral
mapping mode in which a selected IRS slit is moved in a raster pattern,
stopping to obtain one or more exposures at each position, to map out
extended regions.  Since the IRS is a diffraction limited spectrograph
which delivers a PSF varying in size by roughly a factor of 5 over the
full wavelength range, this mode is very useful for obtaining extended
source full-coverage spectrophotometry free from strong aperture bias.
Spectral mapping also offers the significant advantage of preserving
high efficiency for spectroscopic followup of individual point sources,
while providing the additional capability of very deep, spatially
resolved spectroscopy of large, low surface brightness sources, all in a
single instrument without moving parts.

Here we describe an approach to reconstructing full spectral cubes from
scanned or rastered spectroscopic observations using a single slit.  In
\S\,\ref{sec:spectr-cube-reconstr}, we outline a custom algorithm for
cube reconstruction, consider issues of extended source flux calibration
in \S\,\ref{sec:extended-source-flux}, and finally describe \cbsm, a
tool implementing these reconstruction techniques for IRS spectral maps
in \S\,\ref{sec:cubism}.

\section{Spectral Cube Reconstruction}
\label{sec:spectr-cube-reconstr}

The reconstruction of three-dimensional spectral cubes from a suitably
obtained set of overlapping input spectral images (each potentially
consisting of multiple spectral orders) differs fundamentally from image
mosaicing.  The input images carry mixed spatial and spectral
information, the relative spatial offsets among the input spectra cannot
be readily computed from the mapped data themselves, and ensuring flux
conservation requires a dedicated approach.  Here we outline the general
properties of an algorithm for spectral cube reconstruction, discussing
more detail of a specific implementation, along with analysis of the
constructed cube, in the next section.

\subsection{Resampling Noise}
\label{sec:resampling-noise}

\begin{figure}
\plotone{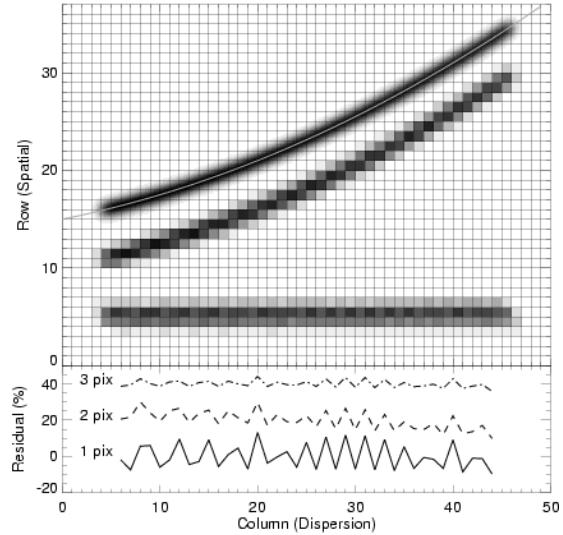}
\caption{An illustration of resampling noise.  In the top panel, above,
  an ideal, uniform, curved spectral order, with center indicated by the
  gray line, is formed from a flat spectrum source with a Gaussian PSF
  of constant width FWHM=1.5 pixels.  It is sampled by the pixel
  elements at middle, and resampled to straighten the order using linear
  interpolation, centered precisely on the known order center, below.
  Extractions of three different pixel widths (offset for clarity) are
  given in the bottom panel, demonstrating the purely sample-based noise
  induced by resampling.}
\label{fig:resamp_noise}
\end{figure}

Modern spectrographs, in particular those which are cross-dispersed,
rarely form spectral images in which the spatial and spectral axes are
aligned with detector rows and columns.  Curved orders which cross many
detector rows are common.  A simple method to deal with these curved
order data involves first resampling them, traditionally with a
flux-conserving form of linear interpolation, to create a straightened
spectral image.  After this step, simple summation can be used to
extract the final spectrum.

Optimal forms of spectral extraction which employ weighted sums to
minimize signal degradation \citep[e.g.][]{Horne1986} could also be
applied to such a straightened image.  In practice, however, an
unavoidable difficulty enters with this approach when the spatial
profile of an unresolved source distributed along the slit is not well
sampled.  As discussed by \citet{Mukai1990}, \emph{resampling noise}
results from the initial straightening step, which is retained and
amplified by all subsequent processing.  Figure~\ref{fig:resamp_noise}
illustrates this effect using a single, idealized curved spectrograph
order formed from a flat spectrum source with a constant Gaussian point
spread function (FWHM=1.5 pixels).  This continuous profile is then
observed with the detector pixel grid (all assumed to have flat and
equivalent response), and then straightened by resampling using linear
interpolation centered precisely on the known order center position.  In
spectra extracted from the resampled order data, periodic resampling
noise of up to 20\% is induced in this example, varying in magnitude
with the size of the extraction region.  The period depends on the rate
at which the order center crosses row boundaries.  Though this noise is
mitigated at larger extraction sizes, this is no consolation when
performing further operations on the resampled spectra, such as
estimating the row by row slit profile for optimal extraction, as
\citeauthor{Mukai1990} considered, or remapping the spatially resolved
pixels along the slit to a common sky coordinate system, as considered
here.

It should be stressed that resampling noise is fully independent from
detector or other instrumental properties, or the form of the point
spread function, but is rather an inherent sampling limitation:
information on the distribution of flux along the slit has been
irrecoverably lost.  As an example, consider a point spread function
(PSF) which is much smaller than the size of a pixel.  In an observed
spectral image, you cannot distinguish between the spectrum of a point
source having been placed in the center of a pixel, or near its edge.
If interpolation is performed to shift this spectrum down by, e.g., one
half pixel, in both cases, two equal interpolated pixels result.  Real
spectrographs present varying and imperfectly determined order
positions, both of which can compound this issue.  Better sampling
alleviates the effect, but it persists at the few percent level even at
the typical sampling of 2 pixels per FHWM.  In diffraction-limited
spectrographs, the PSF sampling can change appreciably from one
wavelength end to the other.  The IRS spectrograph modules
\citep{Houck2004a} were designed to sample the PSF at the long end of
each module's wavelength range, and thus are significantly undersampled
at shorter wavelengths.

To mitigate the effects of resampling noise, it is vital to avoid
additional interpolation steps which can introduce it.  For the
construction of spectral cubes, two opportunities for interpolation can
be avoided: initial straightening of the curved order data, and
remapping of the rastered set of input spectra onto an output cube grid.
Although the information lost by an undersampled spectrum cannot readily
be recovered without imposing prior knowledge on the form of the slit
profile, avoiding unnecessary interpolation --- especially early in the
reduction sequence --- can minimize the resulting sampling noise.  In
typical spectral reduction algorithms employing multiple interpolations
steps, resampling noise can be reduced by $\sim$10 percent using the
methods outlined here, strongly dependent on the signal to noise of the
observations, and intrinsic sampling of the spectral instrument.

\subsection{Spatial vs. Spectral Degeneracy}
\label{sec:spatial-vs.-spectral}

\begin{figure*}
\plotone{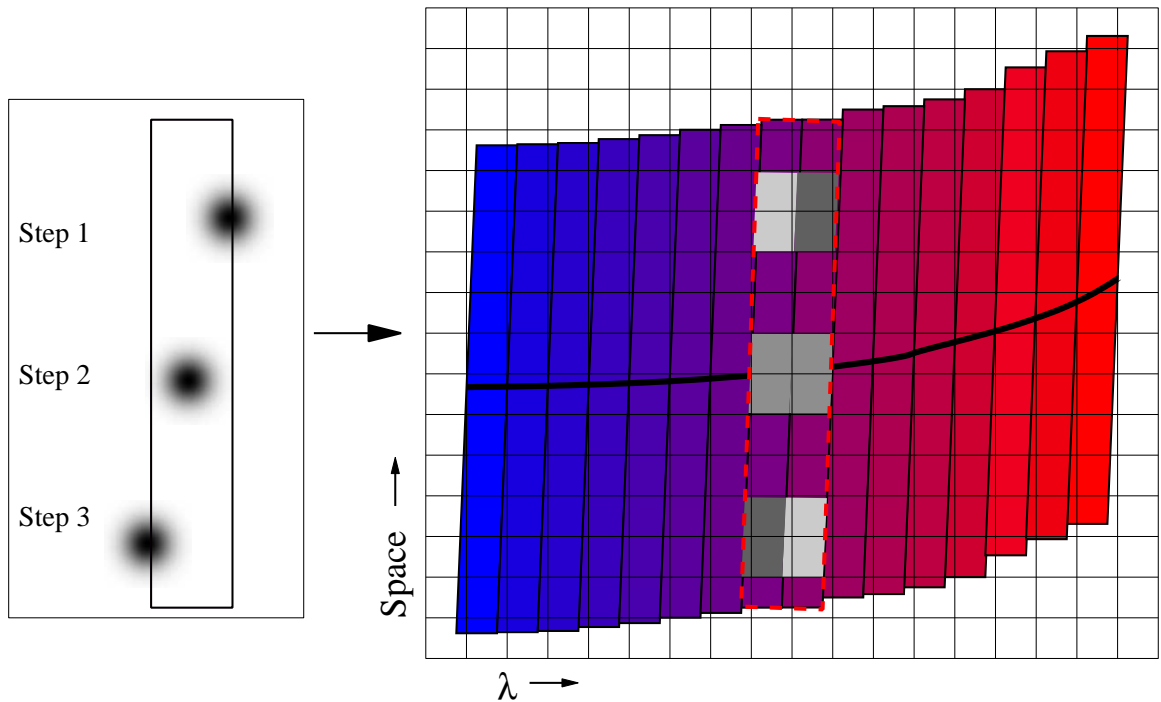}
\caption{An illustration the spectral/spatial degeneracy in sampled
  slits.  A monochromatic point source with PSF smaller than the slit
  width of two pixels is mapped in three steps perpendicular to the slit
  (offset along the slit for clarity).  In the resulting spectral
  images, at right, adjacent pixels in the ``wavelength'' direction can
  be interpreted as adjacent spatially or spectrally.  Single pixel
  width ``pseudo-rectangles'', each of which bounds the contributions
  from a single wavelength, are shown in color.}
\label{fig:spectral_spatial}
\end{figure*}

Creating the three dimensional intensity cube $S(\theta,\phi,\lambda)$
from a series of two-dimensional spectra is fundamentally limited by an
implicit degeneracy in the information present in a spectral image.  As
illustrated in Fig.~\ref{fig:spectral_spatial}, when spectrally mapping
with PSF size comparable to or smaller than the slit width, an ambiguity
exists in the interpretation of pixels adjacent in the dispersion
direction.  Given pixel $p_{ij}$ in a single two-dimensional spectral
image, which maps to the spectral cube location
$S(\theta_{kl},\phi_{kl},\lambda_m)$, the adjacent pixel in the
dispersion direction $p_{i+1,j}$ could be considered to map either to
the same wavelength at adjacent spatial positions within the cube,
$S(\theta_{k+1,l},\phi_{k+1,l},\lambda_m)$, or to the same spatial
position at different wavelengths,
$S(\theta_{kl},\phi_{kl},\lambda_{m+1})$.

In general, both spectral and spatial adjacency are encoded in the slit
image.  When the PSF is large, very little spatial information remains.
When the PSF is small, spatial and spectral information are
irrecoverably convolved together within the slit.  In the latter case, a
single point source with two adjacent spectral lines just blended at the
instrument's spectral resolving power will appear identical to the
spectrum of a source elongated along the slit width, emitting a single
spectral line.  Many image slicing integral field spectrograph designs
mitigate (or hide) this problem by sampling the slit width using only a
single pixel \citep[e.g.][]{Poglitsch2004}, though without methods to
scan wavelength at sub-resolution, this can adversely impact the
reliability of unresolved spectral line measurements.  Addressing this
degeneracy requires some assumption, with three approaches possible:

\begin{enumerate}
\item Assume no spatial information is available in the slit.  Adjacent
  pixels in the dispersion direction are placed into adjacent wavelength
  planes in the spectral cube, and all spatial information is provided
  by the mapping observations.
\item Assume equal spatial and spectral information is available in the
  slit.  Adjacent pixels in the slit are distributed equally into the
  spectral cube within individual wavelength planes (``alongside''), and
  at adjacent wavelengths (``beneath'').
\item Assume a varying mix of spatial and spectral information, which
  can be restated as a mixing angle with respect to planes of constant
  wavelength, varying from 45 degrees (equally spatial + spectral) to 90
  degrees (fully spectral).
\end{enumerate}

In this work, we have adopted the first assumption, though others may be
advantageous in cases of extreme undersampling of the slit.

\subsection{Remapping the Spectra}
\label{sec:spectral-remapping}

\begin{figure*}
\plotone{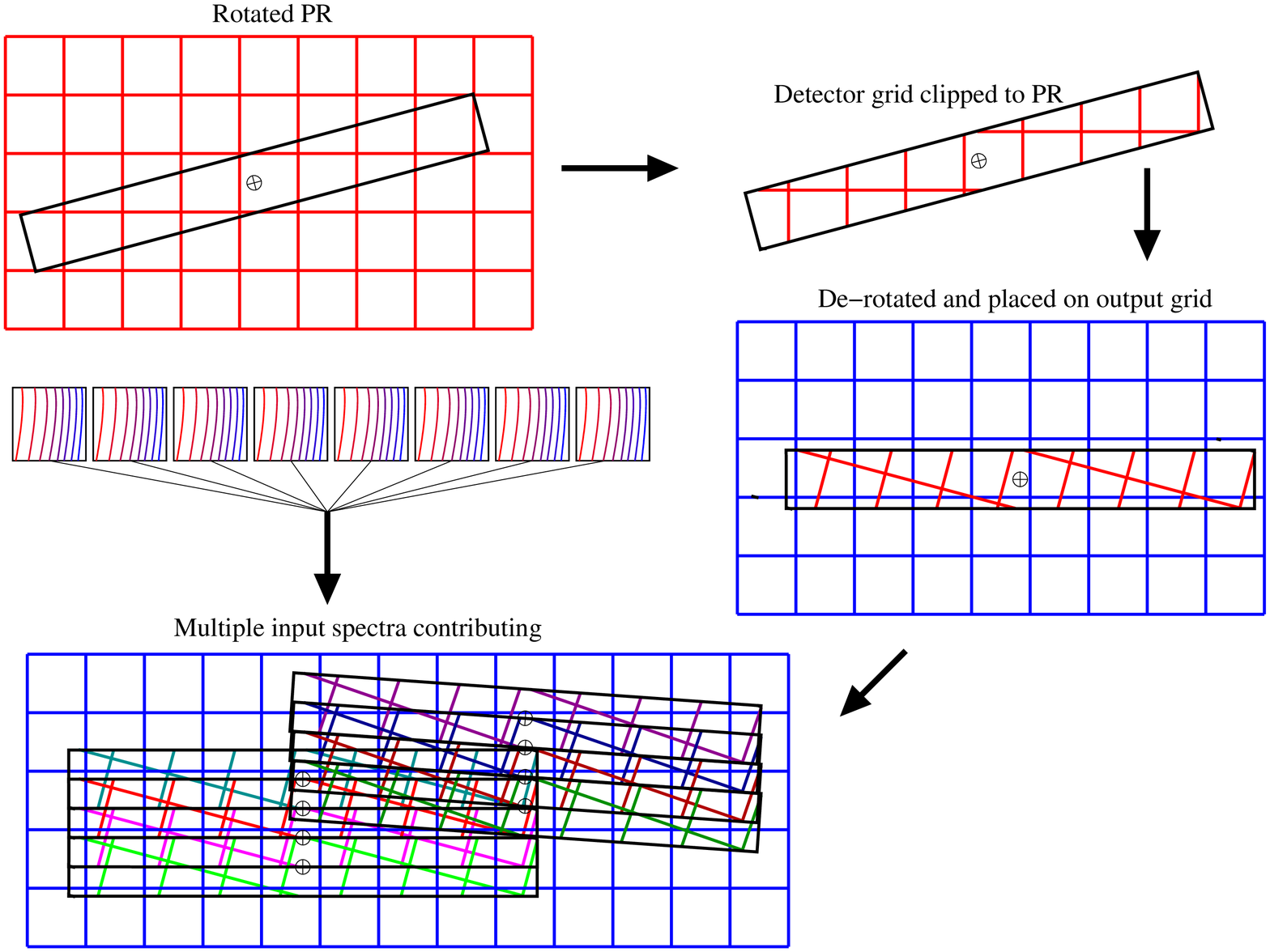}
\caption{An illustration of the two-pass clipping algorithm.  From
  top-left: a single pseudo-rectangle (corresponding to a unique
  wavelength), shown overlain on the input spectral image (red) detector
  grid, is clipped, rotated and offset into the output cube (blue) pixel
  grid at the appropriate wavelength plane.  Multiple overlapping sets
  of input spectra are accumulated in the same fashion.  A contributing
  set of input spectra at a slightly different rotation angle is also
  shown.}
\label{fig:pr_alg}
\end{figure*}

The reconstruction of a full spectral cube $S(\theta,\phi,\lambda)$,
with two spatial ($\theta$,$\phi$) and one spectral ($\lambda$)
dimensions from a set of two-dimensional spectral images which have been
obtained by mapping a given region of the sky could, in principle, be
accomplished in a number of ways.  Pixels from the spectra could be
remapped onto a cube grid, and interpolated.  Conversely, pixels in an
output cube could be reverse mapped into the original spectral image
set, and a weighted average of interpolated values drawn.  In order to
minimize the effects of resampling noise and to ensure flux
conservation, however, a method which avoids unnecessary interpolation
steps is preferable.  

Such a method is available in direct clipped polygon resampling, in
which the contributing weights of individual input pixels are computed
by their geometric overlap with an input mask and output grid.  Polygon
clipping is a mature technique in the field of computer graphics.  For
the purpose of spectral resampling, simple clipping algorithms, e.g.
\citet{Sutherland1974}, suffice.

Since spectral orders are commonly curved, and the orientation of the
spatial and spectral dimensions can vary at each point along the order,
a fundamental element in this algorithm is a means of separating input
pixels by wavelength.  For this purpose, an abutting set of
``pseudo-rectangles'' (PRs), each of which outlines the region of
influence of a single wavelength, can be constructed to cover the curved
order.  The center of each PR is placed at the precise order center, as
illustrated in Fig.~\ref{fig:spectral_spatial}, and they may be angled
to follow the tilt of spectral lines (common in particular in
cross-dispersed spectrograph designs), which itself may vary along the
order.  In this scheme, an individual input pixel can (and often does)
contribute to multiple wavelengths.  The exact method of constructing
the set of PRs is not critical, as long as the central position along
the slit is well determined and matched by the PR centers, there are no
gaps in coverage, and the pixel sampling of the underlying detector
device is respected (which in general means the width of the PR should
be one pixel).

With the set of PRs in hand, the cube reconstruction algorithm consists
of the following steps:

\begin{enumerate}
\item Clip each pseudo-rectangle against the input pixel grid for each
  of $N$ input spectral images $I$ in a given mapping sequence.

\item Rotate and translate the obtained sets of clipped partial pixel
  fragments (grouped by image $I$ and wavelength $\lambda$) into the
  output cube grid, according to position and orientation within the
  map.

\item Clip the offset pixel fragments again against the output (cube)
  grid.

\item Accumulate the overlap area-weighted sum of contributing pixels in
  each cube pixel (and, optionally, a parallel uncertainty cube).
\end{enumerate}

\noindent The final area-weighted sum can be expressed:

\begin{equation}
  \label{eq:1}
  S(\theta_{ij},\phi_{ij},\lambda_m)=\frac{\sum_{k=1}^N\sum_{xy\,\in
    \mathrm{PR}_m}\delta_{xyijm} m_{xy}^{(k)}I_{xy}^{(k)}}{\sum_{k=1}^N\sum_{xy\,\in
    \mathrm{PR}_m}m_{xy}^{(k)}\delta_{xyijm}}
\end{equation}

\noindent where $I_{xy}^{(k)}$ is the data value at pixel $[x,y]$ of the
$k$th input spectral image, $m_{xy}^{(k)}$ is a per-image mask used to
disable input pixels, and $\delta_{xyijm}$ is the area of overlap in
cube pixel $[i,j,m]$ from input pixel $[x,y]$.  The overlap area results
from two distinct rounds of polygon clipping: 1) clipping the
pseudo-rectangle $m$ (corresponding to wavelength $\lambda_m$) against
the input pixel $[x,y]$, and 2) clipping the resulting ``pixel
fragment'' (which can be a full pixel or a polygonal portion of a
pixel), suitably rotated and offset, against the output cube pixel
$[i,j]$.  A schematic of the two-phase clipping approach is shown in
Fig.~\ref{fig:pr_alg}.

The core of the cube reconstruction algorithm is the computation of the
$\delta$ pixel overlap weights.  Their calculation depends on accurate
positioning of the pseudo-rectangles on the input spectrum, including
any line tilt at that wavelength, and the absolute position of the slit
center (and thus PR center) as well as the rotation of the slit in the
output coordinate grid.  Errors in positioning the PR along the order in
the input spectra or spatially within the output grid will degrade the
achieved spectral and/or spatial resolution, as well as increase
correlation of noise in adjacent pixels.  An individual PR may have both
an effective rotation induced by instrumental line tilt in the spectral
image, and a true rotation arising from differential slit position angle
at the time of observation with respect to other spectral images in the
mapping set.  Only the latter should be used to rotate the set of pixel
fragments into the output grid (see Fig.~\ref{fig:pr_alg}).

If the spectrograph produces multiple orders from a single slit
(e.g. using cross-dispersion), the algorithm of Eq.~\ref{eq:1} can be
extended to permit separated groups of input pixels in the wavelength
overlap region between orders to contribute to the same output plane.
The wavelength sampling of the output cube in the overlap region between
orders is chosen to accommodate the maximum spectral resolution present
there, and the $\delta$ weighting terms are split linearly between the
bracketing output planes, according to their proximity to the output
wavelength, e.g. $\delta\rightarrow [f\delta, (1-f)\delta]$, where
$f=(\lambda_2-\lambda_0)/(\lambda_2-\lambda_1)$, for an input wavelength
$\lambda_0$, and its bracketing output wavelength pair
$[\lambda_1,\lambda_2]$.  This ensures flux conservation even for
wavelengths sampled in multiple orders.

If accurate error estimates are available for individual pixels in the
input spectral images, they can be propogated through Eq.~\ref{eq:1}
using identical weight factors to build a parallel variance cube:

\begin{equation}
  \label{eq:2}
  \sigma^2(\theta_{ij},\phi_{ij},\lambda_m)=\frac{\sum_{k=1}^N\sum_{xy\,\in
    \mathrm{PR}_m}\delta_{xyijm} m_{xy}^{(k)}V_{xy}^{(k)}}{\sum_{k=1}^N\sum_{xy\,\in
    \mathrm{PR}_m}m_{xy}^{(k)}\delta_{xyijm}}
\end{equation}

\noindent where $V_{xy}^{(k)}$ is the variance per input pixel.  This
algorithm does introduce noise correlations between adjacent pixels (as
would interpolative approaches), but in practice these are minimized by
employing a high level of pixel redundancy in filled spectral maps, with
often a dozen or more input pixels contributing to a single output
pixel.  Such pixel redundancy occurs naturally for overlapping spectral
orders, and is greatly increased by constructing mapping observations in
which single positions on the sky are sampled by several different parts
of the slit (along both its width and length).

The polygon-clipping method discussed here bears a superficial
similarity to the DRIZZLE algorithm of \citet{Fruchter2002}, but differs
in that it employs exact pixel overlap calculations, two levels of pixel
clipping (at the pseudo-rectangle stage, and again into the output
cube), and does not shrink pixels or pixel framents via a ``drop
factor''.  Its advantage over related methods to remap curved spectral
data is primarily in avoiding unnecessary interpolation steps, which can
introduce resampling noise early in the reduction process, as well as
enforcing surface brightness conservation implicitly.  Another advantage
is the ability to track individual contributions from spectral pixels in
the source images to the final cube, which can be used readily to reject
discrepant pixels.

\subsection{Pixel Rejection and Masking}
\label{sec:pixel-reject-mask}

The algorithm introduced offers significant statistical power for
identifying and rejecting transient or persistent bad pixels.  Depending
on the sampling redundancy at each pixel within the spectral cube, sigma
trimming methods can be used effectively.  Since a given input pixel
$[x,y]$ will appear many times in a large spectral cube, and since each
output pixel can have many (dozens or more, depending on map layout)
contributing input pixels, statistics can be accumulated to flag outlier
pixels.  Various schemes can be employed, but a sigma threshold,
normalized to an effective measure of the natural spread in the data ---
the median absolute deviation from the median --- proves effective. To
identify persistent deviant pixels in all input spectra, it suffices to
demand that a given input pixel (at a specific $[x,y]$) is flagged at
least some fraction of the times it appears in the output (e.g. 50\%).
See \S\,\ref{sec:bad-pixels} for a specific implementation of this
approach.

\section{Extended Source Flux Calibration}
\label{sec:extended-source-flux}

Accurate flux calibration of extended sources has always proved
challenging for slit spectrographs, since uniform, beam-filling
astrophysical reference sources of known flux intensity (in MJy/sr, for
instance) as a function of wavelength are not generally available.
Instead, point sources (typically stars) with well-measured or modeled
flux distributions are used to provide spectrophotometric reference.
The flux-calibration of point sources is then, ideally at least, trivial
--- as long as all science targets are unresolved, and are placed in
exactly the same position within the slit as the reference calibration
sources, all details of beam acceptance profile and diffractive losses,
which can vary significantly with wavelength and position along the
slit, precisely cancel out, and the ratio of unbiased spectra between
target and calibration reference will form an accurate physical flux
ratio with the model flux.  This ideal situation, though approachable
for point sources, is by definition not possible for extended sources,
light from which enters the slit at all allowable angles, modulated by
the source's angular distribution within the slit.  For this reason,
some care must be taken to obtain accurate spectrophotometric
calibration of extended sources, and in particular spectral cubes
reconstructed from mapping observations of these sources.

There are three main issues with extended source calibration based on
point source spectrophotometric references, which will be discussed in
order of their ease of treatment.

\subsection{Aperture Correction}
\label{sec:apert-loss-corr}

In imaging point source photometry, a crucial issue for calibrating
extended sources is the aperture over which the point source flux is
measured, typically stated as a radius in pixels.  Assuming all
unresolved sources are measured with the same aperture radius, the
derived fluxes do not depend on the aperture size.  However, since such
an aperture encircles only a fraction of the total flux of a source, an
\emph{aperture correction} must be applied to derive accurate extended
source calibrations from known photometric reference stars
\citep[e.g.][]{Reach2005}.

An identical issue occurs in slit spectroscopy, with the extraction
aperture width serving the role of the photometric radius.  Small
aperture widths are desirable for faint sources, since they exclude
noisy pixels in the source's profile along the slit.  Larger aperture
widths, however, minimize aperture corrections, since they capture the
bulk of the source flux present in the spectrum.  This tension between
the desire for larger aperture for unbiased flux measurement, and
smaller apertures for high signal-to-noise flux recovery has lead to PSF
fitting methods \citep[e.g.][]{Stetson1987}, and their approximate
analog in spectroscopic applications, optimal extraction
\citep[e.g.][]{Horne1986}.

Neither PSF fitting nor optimal extraction are applicable to extended
sources, since they rely on the uniform distribution of source flux on
the detector or in the slit.  The first step of accurate extended source
flux calibration is, therefore, forming an estimate of the aperture
losses as a function of wavelength (the \emph{aperture loss correction
  function} --- ALCF).  This function can be applied to the
normally-extracted spectrum of spectrophotometric reference stars to
estimate the full, undiminished flux at each wavelength in the limit of
an infinite aperture.  For long slits, the form and magnitude of the
ALCF can often be obtained trivially using bright flux reference stars
with wide extraction apertures.

\subsection{Slit Loss Correction}
\label{sec:slit-loss-correction}

A slit has finite dimensions, and therefore accepts only a portion of
the flux of a point source imaged onto it.  In diffraction limited
spectrographs, such as the IRS, this slit throughput changes appreciably
with wavelength, as the PSF delivered by the telescope grows in size by
diffraction.  For calibrating point source spectra, this does not
present a difficulty; as long as all point sources, including the flux
references, are well-centered in the slit, the slit throughput function
divides out.  For extended source, however, a calibration system tied to
reference point source observations will necessarily include an implicit
correction for the fraction of the point source flux which the slit
admits.  In the limit of a perfectly uniform, slit-filling extended
source --- which can be conceptualized as an infinitely dense grid of
point sources with vanishing flux --- the diffractive losses \emph{out
  of} the slit are exactly offset by the diffractive gains \emph{into}
the slit from sources beyond its geometric boundary.

\begin{figure}
\plotone{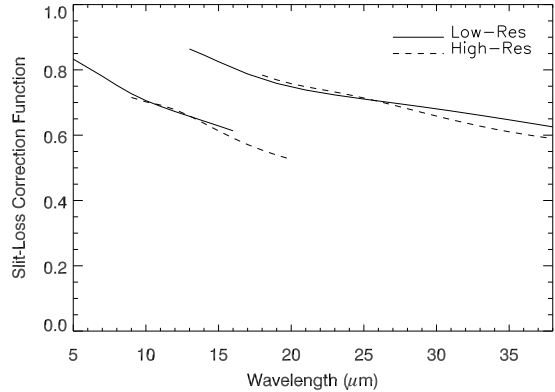}
\caption{The slit-loss correction function for the four IRS modules.}
\label{fig:slcf}
\end{figure}

In this limit, then, a \emph{slit loss correction function} (SLCF),
which is simply the fraction of the flux of a well-centered point source
admitted by the slit as a function of wavelength, must be used to
correct extended source fluxes.  Without this correction, extended
source fluxes are overestimated.  Obviously, the actual magnitude of the
true SLCF depends on the observed source flux distribution, which can be
intermediate between unresolved and fully uniform.  This complicates its
use, but in practice, depending on the size of the slit, this detail can
often be ignored, and a single correction function can be employed for
all extended sources.  Estimates of the SLCF can be obtained by
convolving models of the PSF (e.g. STinyTim for the Spitzer telescope)
with the approximate slit dimensions.  An example of such a set of
SLCF's for the four spectrograph modules of Spitzer's IRS is shown in
Fig.~\ref{fig:slcf}.

\subsection{Slit Beam Profile}
\label{sec:beam-profile}

Though a slit can be formed with precise geometric boundaries, the solid
angle it subtends on the sky is broadened due to diffraction.  Just as
for radio dish measurements, a slit has some beam profile,
$B(\theta,\phi)$, which can change with wavelength.  Estimating this
beam profile function can be very difficult.  \citet{Salama2000}
describe a method for de-convolving the spectrograph beams aboard the
Infrared Space Observatory using a series of spectra obtained by
rastering a point source around the input slit.  Such a method obviously
relies heavily on models of the diffraction of the combined telescope
and spectrograph optical path, but in principle can be used to recover
$B$.

When attempting to test the reliability of a given slit loss correction,
for instance by comparing synthetic photometry from the spectra to real
photometry over a common wavelength range (e.g. 24\um\ Spitzer
photometry) there is a strong degeneracy with the effective solid angle
subtended by the slits, $\Omega_\mathrm{eff}=\int B(\theta,\phi) d\theta
d\phi$, such that the SLCF can be scaled up or down, with compensating
changes to the inferred beam.  If the slit is uniform, and unless both
corrections are separately known, it is useful to collapse all
information of the beam profile to a single number, the effective solid
angle per pixel, and derive this number by comparing photometry with
spectrophotometry in a sample of extended source measurements.  This
effective solid angle per pixel should be approximately equal to the
slit width times the pixel scale.

\section{\textsc{Cubism}}
\label{sec:cubism}

An implementation of the cube reconstruction algorithm presented in
\S\,\ref{sec:spectr-cube-reconstr} for spectral mapping observations
using Spitzer's IRS spectrograph is available in \cbsm, the CUbe Builder
for IRS Spectral Mapping\footnote{Other tools are available for the
  processing of single, staring mode IRS observations --- e.g. SMART
  \citep{Higdon2004} and SPICE.}.  \cbsm\ is written in the Interactive
Data Language (IDL), and is designed to combine sets of two-dimensional
spectral images from IRS mapping observations into single 3D spectral
cubes, with two spatial and one spectral dimension.  A variety of cube
analysis tools are also available, including arbitrary extractions and
map creation (e.g. a continuum-subtracted line image).

\cbsm, and a manual detailing its use, are available from the Spitzer
Science
Center\footnote{\url{http://ssc.spitzer.caltech.edu/archanaly/contributed/cubism}}.

\subsection{Inputs and Calibration}

\cbsm\ uses by default the raw two-dimensional basic calibrated data
(BCD) spectral images, which have been processed by the IRS pipeline
(flat-fielded, corrected for straylight and certain detector anomalies,
etc.).  Versions of the data without flat-fielding or stray-light
correction can optionally be used for testing purposes.  It applies the
twin-clipping reprojection algorithm of
\S\,\ref{sec:spectr-cube-reconstr}, using an input set of calibration
data derived primarily from SSC calibration products.  A set of
pseudo-rectangles for each order (see \S\,\ref{sec:spectral-remapping}),
which delineate the order position and wavelength solution, are
constructed directly from the order position and line tilt calibrations.
The standard \emph{WAVSAMP} calibration files provide a similar set of
PRs, but do not follow the detector's pixel sampling as required to
minimize noise correlation, and are therefore not employed.  Extended
source SLCF and ALCF flux corrections are applied by default (see
\S\,\ref{sec:extended-source-flux}), the latter by employing a special
form of the standard \emph{FLUXCON} flux calibration input.

The effective solid angle per pixel (the single free parameter in the
extended source flux calibration methods discussed in
\S\,\ref{sec:extended-source-flux}) is calibrated by comparing
integrated spectro-photometry over $\sim$arcmin$^2$ areas to corrected
photometry in MIPS, IRAC, IRS Blue Peak-up, and ISOCAM data in bright
galaxies from the SINGS Legacy survey \citep{Kennicutt2003}.  In all
cases, these solid angle terms agree with the measured slit width within
10\%.

\subsection{Cube Construction}

\cbsm\ constructs an output cube grid which bounds the mapped region,
with default pixel size identical to the sampling in the relevant
spectrograph module (ranging from 1\farcs85 to 5\farcs08).

Individual spectra are offset into the output grid according to the
requested or reconstructed positions of the slit center during the
exposure.  The latter, based on contemporaneous telescope attitude
reconstruction from gyroscopes and star tracker, yield more stable
results.  The limited spatial information available in any individual
spectral image makes cross-correlation alignment untenable, such that
accurate positional accuracy (below the level of the pixel sampling) is
crucial for this technique.  Resultant positional errors, compared to
fixed astrometric systems (e.g. 2MASS) are $\sim$1\arcsec.

Each of the six sub-slits --- two each for low-resolution modules
long-low (LL, 15--37\um) and short-low (SL, 5--15\um), along with two
high resolution modules short-high (SH, 10--20\um) and long-high (LH,
20--38\um) --- are treated independently.  A single cube can contain
data for only a single sub-slit, though for the low-resolution modules,
it is unimportant how the data were targeted (e.g. the second order of
SL can be used from BCDs in which the first order was targeted,
etc.). In the latter case, the frame table which defines absolute
positions of all Spitzer apertures is consulted to calculate the
position of the offset sub-slit.  Consistent with measurements, spatial
distortion along the slits due to projection distortion or other optical
effects is assumed negligible.  In each of the high-resolution modules,
ten orders are formed by cross-dispersion from the single entrance slit.
The cubes are created by merging the order data in the overlap wavelenth
region, as described in \S\,\ref{sec:spectral-remapping}.

The fractional contribution of all input pixels contributing to a given
output cube pixel are computed, and are collectively called the cube
\emph{accounts}.  These can be stored and reread, and permit high level
statistical methods to be applied to the cube data, for example to
exclude outliers (see \S\,\ref{sec:bad-pixels}).

For speed, the core clipping functions, which implement a
special-purpose reentrant form of the Sutherland-Hodgman polygon
clipping algorithm \citep{Sutherland1974}, is written in C and
auto-compiled.  If auto-compilation fails, an IDL-native version will be
used as fallback (at a significant speed penalty).

\cbsm\ was extensively tested on simulated spectral mapping data to
validate the algorithm and ensure flux conservation.

\subsection{Uncertainty}

Full error cubes are built alongside the data cubes by standard error
propagation of Eq.~\ref{eq:1}, using, for the input uncertainty
$\sigma_{I_{xy}}$, the BCD-level uncertainty estimates produced by the
IRS pipeline from deviations of the fitted ramp slope fits for each
pixel.  These uncertainties, which are statistical ramp uncertainties
only, and neglect other errors in calibration and later processing
steps, are used to provide error estimates for extracted spectra and
constructed line or continuum maps.

\subsection{Background}
\label{sec:background}

Both to remove the astrophysical foreground and backgrounds (primarily
zodiacal and cirrus), and to mitigate the effects of time varying warm
or \emph{rogue} pixels, it is advantageous to subtract near in time
background frames from each input data record at the 2D level.  Suitable
background frames (which are free from source contamination in the
targeted slit), can be taken from the observations directly (e.g. for
smaller targets), from dedicated offset observations, or, if neither is
available, from records obtained from the archive at close times (within
a few days) and ecliptic latitude ($|b-b_0|\lesssim10\degr$).  If
necessary, additional correction can be made by subtracting a 1D
spectrum (e.g. extracted from a dark region within the cube) ---
equivalent to removing a constant value from each cube plane.

\subsection{Bad Pixels}
\label{sec:bad-pixels}

The bad pixel masks --- $m_{xy}^{(k)}$ of Eq.~\ref{eq:1} --- are derived
from three sources.  Permanently marked non-responsive pixels from the
\emph{PMASK} are combined with pixels flagged during the exposure for
saturation (zero or one ramp sample valid), or flat-field difficulties,
which are reported in the \emph{BMASK} input frames (automatically
loaded alongside the BCDs).  Additional user bad pixels can be marked
either globally (such that pixel $[x,y]$ is disabled in all input
spectra), or for individual input records.  Both types of bad pixels can
be generated automatically, by computing outlier statistics in the
accounts information built up during the cube generation (the full
record of which input pixels contributed to which output pixels, and
with what fractional weight).  Two parameters, $r_\sigma$, the
sigma-trim threshold, and $f_{min}$, the minimum fraction, control
automatic bad pixel detection.  For each output cube pixel, the
contributing pixel residuals (either with or without subtracted
background) are tested against an estimate of the contributing pixel
deviation $\sigma$ (computed either from a true variance or, for small
number contributing, the median absolute deviation from the median
value).  If a pixel is an outlier,
i.e. $|I_{xy}-\mathrm{median}(I_{xy\in ijm})|\ge r_\sigma \sigma$, it is
tentatively marked bad.  If the same pixel is so marked at least a
fraction $f_{min}$ of the times it appears in teh cube, it is flagged.
Significant power at high spatial frequency (point sources, edges)
results in a large natural range of contributing pixel values, such that
some care must be taken to ensure valid data (e.g. strong spectral
lines) are not flagged.

\subsection{Components}

The \cbsm\ interface consists of three main components: the project
window, a multi-purpose viewer, and a spectrum viewer and map creation
interface.  Underlying the interface components is a scriptable
object-oriented data container, serialized to disk using IDL's own save
format (e.g. `cube.cpj').  All components communicate with each other to
present a consistent state.  For example, when loading a new set of bad
pixels in one component, the displayed bad pixels in another are
automatically updated.

\subsubsection{Cubism Project}
\label{sec:cubism-project}

\begin{figure}
\epsscale{1.18}
\plotone{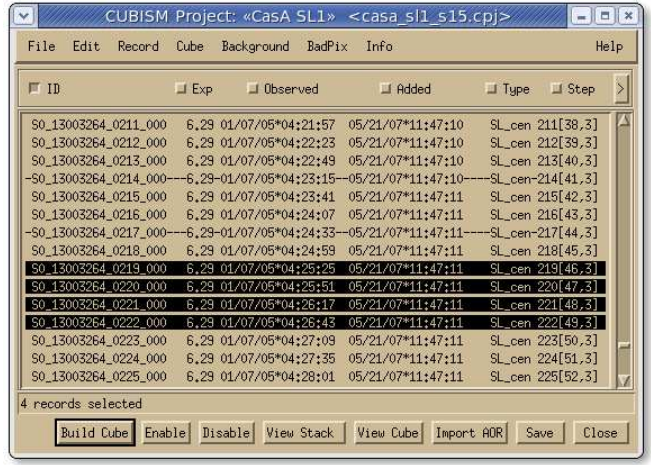}
\epsscale{1.0}
\caption{The \cbsm\ Project window, with several disabled and selected
  records.}
\label{fig:cubism_project}
\end{figure}

All data, calibration inputs, and cube build settings are collected and
operated on from the Cubism Project pane, shown in
Fig.~\ref{fig:cubism_project}.  From this interface, projects can be
loaded and saved, or recovered from disk.  Any number of project windows
can be opened at a time.  Records (individual BCDs) can be added and
removed, disabled from the build, combined into background images in
various ways, sorted, examined for header information, etc.  The cube
build parameters and cube build itself are controlled from the project
component.  The \cbsm\ project interface also allows bad pixels and
background records sets to be loaded, combined, and saved.  An
individual project targets a single slit in a single module.  Separate
visualization images can be loaded, and feedback is provided on the cube
build.  Various optimizations are employed to speed up the cube build
process, including keeping track of the specific cube pixels affected by
changes to bad pixel settings, and re-building only these.

Using the viewer component, records or stacks of records, the assembled
cube, or an image loaded for visualizing the position of the slit at
each step can be displayed.

\subsubsection{CubeView}
\label{sec:cubeview}

\begin{figure*}
\epsscale{1.15}
\plotone{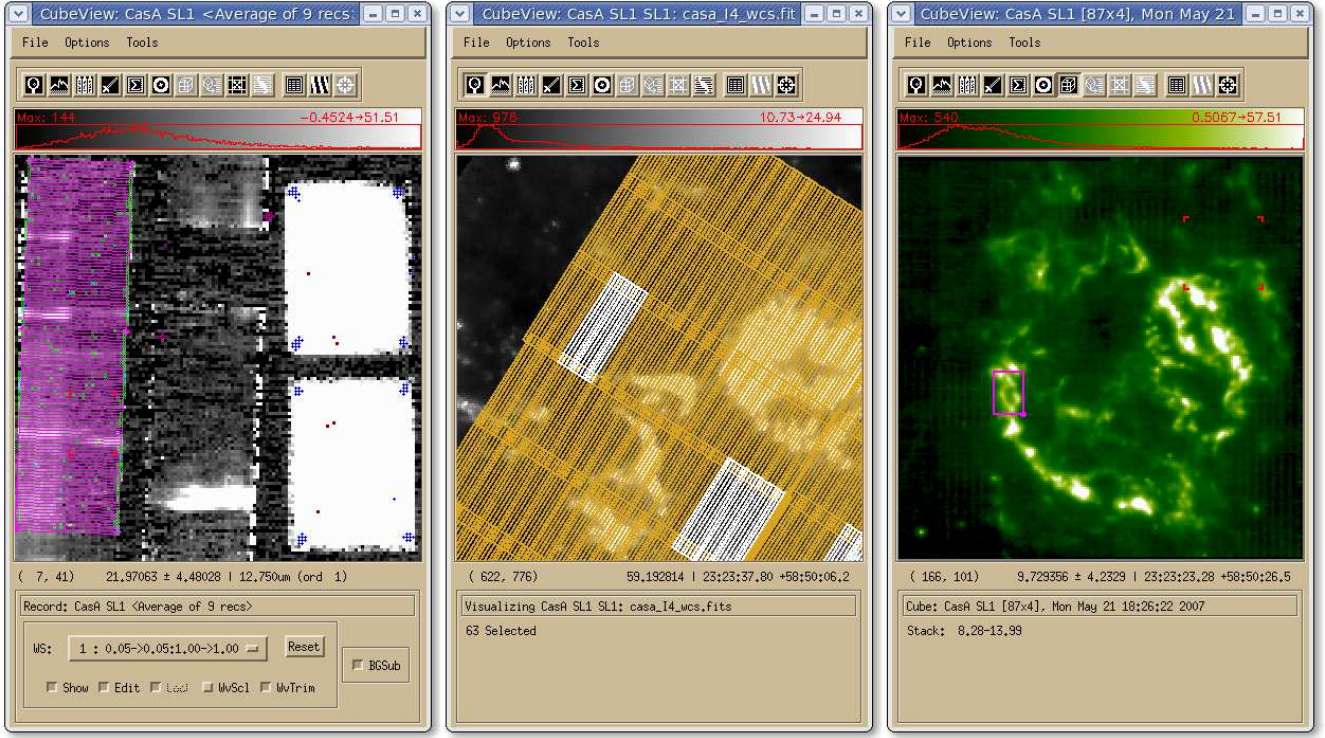}
\caption{The CubeView viewer, in three different configuration, showing
  (from left to right), a stack of BCD records with pseudo-rectangle set
  overlayed and bad pixels displayed, a visualization image with the
  slit positions and selected records shown, and a stack built from the
  cube itself with an extracted region shown.}
\label{fig:cubeview}
\end{figure*}

Individual records, stacks of records, spectral cubes or maps, and
visualization images can all be viewed in a single viewer interface:
CubeView.  CubeView has a number of general and special purpose tools
that are activated as needed.  A single project at a time communicates
with one or more instances of the viewer, which modifies itself based on
the type of data displayed.  For example, if the cube project sends for
view an average stack of BCD records, CubeView provides tools for
toggling background-subtraction, editing the slit-end cutoff positions
of the pseudo-rectangle set, viewing, adding, or removing global and
record level bad pixels, etc.  When viewing full spectral cubes, an
interface to move through the cube planes is provided, and a cube
extraction tool is enabled, to construct rectangular extractions or
matched region extractions based on preexisting spectra.  When viewing
cubes, maps, or visualization images, the WCS coordinates are reported,
whereas with spectral data, the wavelength, order, and any pixel flags
are shown.  When viewing visualization images, individual records can be
selected directly on the overlay (useful for specifying background
records).  General use tools, including histogram scaling, color map
modification, line slice plotting, box statistics, aperture photometry,
a pixel table, and others, are always available.

Multiple CubeView instances can be used simultaneously, or a single
instance reused to view different types of data.  Separate projects
loaded in the same session communicate with their own set of CubeView
tools.  An example of CubeView displaying record spectral data, a
visualization image, and map created from a spectral cube is shown in
Fig.~\ref{fig:cubeview}.

\subsubsection{CubeSpec}
\label{sec:cubespec}

\begin{figure*}
\epsscale{0.63}
\plotone{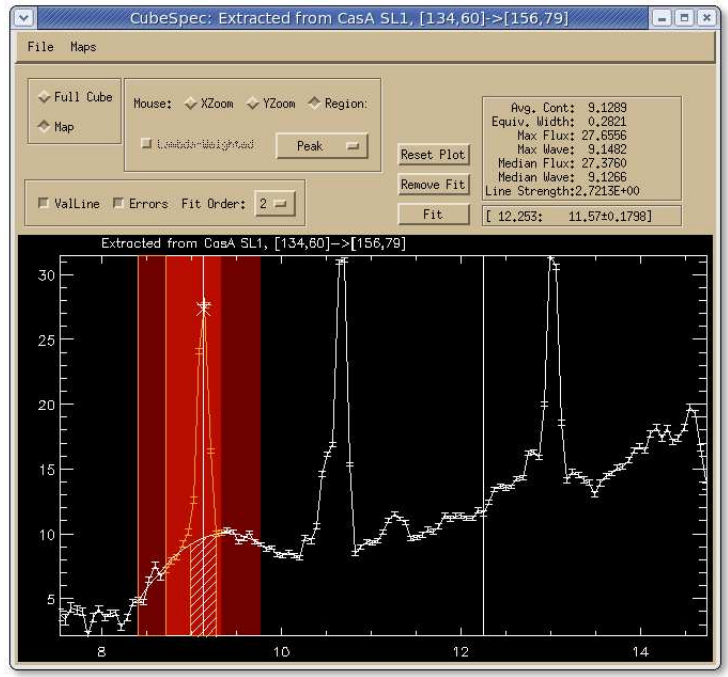}
\caption{The CubeSpec window, displaying an extracted spectrum, and map
  region comprised of one peak and two continuum regions.  The line has
  been fitted.}
\label{fig:cubespec}
\end{figure*}

When an extraction is made from a spectral cube being viewed, the
resulting spectrum is displayed in the CubeSpec tool, shown in
Fig.~\ref{fig:cubespec}.  Map images can be assembled and directly
viewed in the associated CubeView viewer by specifying an arbitrary
number of foreground and continuum wavelength regions over which to
average (e.g. a line region flanked by two continuum regions), with
either wavelength-offset weighted or uniformly weighted averages used to
set the continuum in each foreground plane.  Map sets can be saved and
restored by name, and a custom redshift entered to shift sets
(e.g. specified at rest frame) into the observed frame of a target.  The
tool also provides simple line fitting capabilities, and the ability to
create maps using filter curves or other weighting functions (e.g. a
24\um\ image using the MIPS 24\um\ filter curve).  

\subsection{Outputs}
\label{sec:outputs}

\cbsm\ outputs data products in standard FITS and IPAC Table formats.
Although the working storage format is IDL SAV file, the full spectral
cubes can be exported as FITS, with the third spectral dimension's
wavelengths encoded as a lookup table, following the FITS spectral
coordinate standards of \citet{Greisen2006}.  Maps created from the
cubes are output as standard 2D FITS images, with information encoding
the foreground and background wavelength range(s) used to create them.
When uncertainty cubes have been built, accompanying uncertainty FITS
files are produced alongside the FITS data products.

Spectral extractions are reproduced in the IPAC Table format (an ASCII
format with column delineations, labels, and
units)\footnote{\url{http://irsa.ipac.caltech.edu/applications/DDGEN/Doc/ipac\_tbl.html}},
and include information describing the coordinates of the extraction
rectangle.  All products have associated WCS coordinate systems
attached, with estimated accuracy of 1\arcsec.

\section{Optimized Spectral Mapping}
\label{sec:opt-map}

As for deep image mosaics, the redundancy of pixel sampling is the
foremost determinant of the quality of a produced spectral cube.
Redundancy permits robust statistical rejection of bad pixels, minimzes
the impact of variations in slit throughput or detector response along
the slit, and reduces noise correlation between neighboring output
pixels.

In particular for the IRS, with its time varying rogue pixels, sampling
a given position on the sky with at least 4 different detector pixels is
crucial for achieving high final cube quality (e.g. using the methods of
\S\,\ref{sec:bad-pixels}).  Given the spatial and spectral degeneracy
(\S\,\ref{sec:spatial-vs.-spectral}), it is also critical to step by
one-half the slit width in the dispersion direction.  Wider step spacing
risks gaps or uneven photometric response in the spectral cube, in
particular when a random and unkown spatial dither pattern is added to
the requested pointing positions.  A detailed set of recommendations for
planning and executing spectral maps is provided at the
SSC\footnote{\url{http://ssc.spitzer.caltech.edu/irs/documents/specmap\_bop}}.

\section{Summary}
\label{sec:summary}

\begin{figure}
\plotone{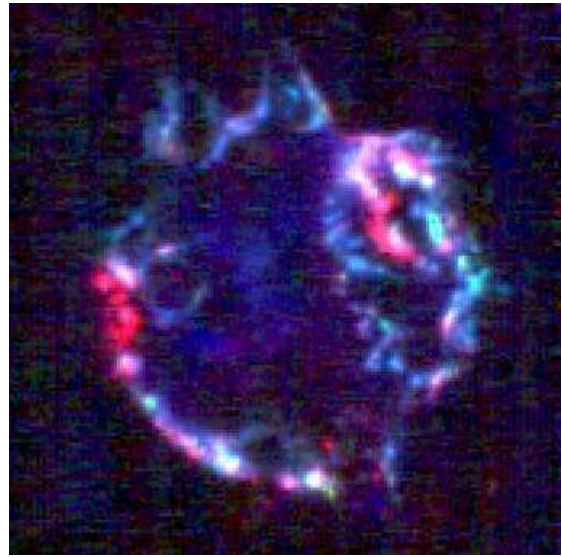}
\caption{An example map created from a SL IRS spectral map (7--15\um) of
  supernova remnant Cas A, in three continuum-subtracted lines: \neII\
  (red), \arIII\ (green), \sIV\ (blue).  Each position in the map has an
  associated full 5--38\um\ spectrum. }
\label{fig:casa}
\end{figure}

In observing regimes with stable astrophysical background and image
characteristics, integral field spectroscopy of extended sources can be
achieved efficiently by rastering a single long slit to cover the target
area with suitable redundancy.  This can significantly reduce instrument
complexity and size, in particular if a long slit mode is already
required for efficient followup of isolated point sources.  These
savings in instrument complexity are traded against increased dependence
on telescope pointing accuracy --- only telescope attitude systems which
can track position at angular scales substantially smaller than the
smallest slit width provide effective platforms for slit mapping
integral field spectroscopy.

A resampling noise minimizing, two-pass polygon clipping algorithm for
the reconstruction of spectral cubes with two spatial and one spectral
dimension from spectral mapping data sets was introduced, and an
implementation of this algorithm for Spitzer IRS spectral mapping
observations, \cbsm, was described.  \cbsm\ has been used for individual
maps comprised of $\sim$9000 individual spectral data frames, for
ultra-deep blind spectral mapping surveys, and for large area mosaics
covering hundreds of square arcmin.  It makes full use of the unparalled
background-limited performance of the IRS spectrograph to provide
sensitive integral field infrared spectrsocopy.  An example custom image
created from three continuum-subtracted lines from a large map of the
supernova remnant Cassiopeia A is shown in Fig. \ref{fig:casa}.

\acknowledgments 

The authors thank the staff of the IRS instrument support team at the
Spitzer Science Center, Caltech --- with special thanks to Phil
Appleton, Carl Grillmair, David Shupe, Pat Morris, Sergio
Fajardo-Acosta, Patrick Ogle, Jim Ingalls, Harry Teplitz, and Patrick
Ogle --- for substantial continuing support and advice regarding all
aspects of telescope control, instrument calibration, pipeline behavior,
and more.  We are also indebted to the IRS instrument team at Cornell
University --- in particular Jim Houck, Vassilis Charmandaris, Jeff Van
Cleve, Keven Uchida, Greg Sloan, Sarah Higdon, and Daniel Devost --- for
much helpful advice, discussions, calibration assistance, and insight
into the instrument.  Michael Cushing provided helpful discussions
regarding aspects of the algorithm design.  Support for this work, part
of the \textit{Spitzer Space Telescope} Legacy Science Program, was
provided by NASA through Contract \#1224769 issued by JPL/Caltech under
contract \#1407.

\bibliographystyle{apj}
\bibliography{general,myref,inprep}

\end{document}